\documentclass[fleqn,usenatbib,usedcolumn]{mnras}
\usepackage[british]{babel}
\usepackage{newtxtext}
\usepackage[slantedGreek]{newtxmath}
\usepackage[T1]{fontenc}
\usepackage{graphicx, mathrsfs}

\title[Energy Losses In Few-body Codes]{Implementing Tidal and Gravitational Wave Energy Losses in Few-body Codes: A Fast and Easy Drag Force Model}

\author[J. Samsing et al.]{
Johan Samsing$^{1,}$\thanks{E-mail: jsamsing@gmail.com},
Nathan W. C. Leigh$^{2,3,4}$,
Alessandro A. Trani$^{5}$
\\
$^{1}$Department of Astrophysical Sciences, Princeton University, Peyton Hall, 4 Ivy Lane, Princeton, NJ 08544, USA\\
$^{2}$Department of Astrophysics, American Museum of Natural History, Central Park West and 79th Street, New York, NY 10024, USA\\
$^{3}$Department of Physics and Astronomy, Stony Brook University, Stony Brook, NY 11794-3800, USA\\
$^{4}$Center for Computational Astrophysics, Flatiron Institute, 162 Fifth Avenue, New York, NY 10010, USA\\
$^{5}$Department of Astronomy, Graduate School of Science, The University of Tokyo, 7-3-1 Hongo, Bunkyo-ku, Tokyo, 113-0033, Japan}

\date{Accepted XXX. Received YYY; in original form ZZZ}

\pubyear{2018}

\begin{document}
\label{firstpage}
\pagerange{\pageref{firstpage}--\pageref{lastpage}}
\maketitle

\begin{abstract}

We present a drag force model for evolving chaotic few-body interactions with the inclusion of orbital
energy losses, such as tidal dissipation and gravitational wave (GW) emission. The main effect from
such losses is the formation of two-body captures, that for compact objects result in GW mergers,
and for stars lead to either compact binaries, mergers or disruptions. Studying the inclusion of energy loss terms in few-body interactions
is therefore likely to be important for modeling and understanding the variety of transients that soon will be observed by current and
upcoming surveys. However, including especially tides in few-body codes has been shown to be technically difficult and computationally heavy,
which has lead to very few systematic tidal studies. In this paper we derive a drag force term
that can be used to model the effects from tidal, as well as other, energy losses in few-body interactions,
if the two-body orbit averaged energy loss is known a priori. This drag force model is very fast to evolve,
and gives results in agreement with other approaches, including the impulsive and affine tide approximations.

\end{abstract}




\section{Introduction}

Transient events, including gravitational wave (GW) mergers \citep{2016PhRvL.116f1102A,
2016PhRvL.116x1103A, 2016PhRvX...6d1015A, 2017PhRvL.118v1101A, 2017PhRvL.119n1101A, 2017PhRvL.119p1101A}, stellar
mergers \citep[e.g.][]{2011A&A...528A.114T}, and stellar tidal disruptions \citep[e.g.,][and references therein]{2016ApJ...823..113P}, are often
the product of a two-body or a dynamical few-body system
loosing orbital energy through one or more dissipative mechanisms. The most important of such mechanisms include energy
dissipation through the the emission of GWs \citep[e.g.][]{Peters:1964bc, Hansen:1972il, 1977ApJ...216..610T}, and orbital energy losses through
tidal excitations \citep[e.g.][]{1977ApJ...213..183P, 1986ApJ...310..176L} and dissipation \citep[e.g.][and references therein]{2014ARA&A..52..171O}. In the
isolated binary problem, these effects will lead to a merger between the two objects within a finite time, and depending on the
stellar types the final binary evolution will either be dominated by GWs \citep[e.g.][]{Peters:1964bc} (if both objects are compact), 
tides \citep{2014ARA&A..52..171O} (if at least one object is a star), or common envelope
evolution \citep[e.g.][]{1976IAUS...73...75P, 1993PASP..105.1373I, 2000ARA&A..38..113T, 2018arXiv180303261M} (if one of the objects evolves to indulge the other).

During chaotic interactions involving three or more objects, the loss or dissipation of orbital energy often results in the formation of
eccentric two-body captures \citep[e.g.][]{1992ApJ...385..604K, 2014ApJ...784...71S, 2016ApJ...823..113P, 2017ApJ...846...36S}.
A capture refers here to a scenario involving a very close approach between two objects with such a small pericenter distance that the energy loss over one orbit is large
enough for the two objects to quickly inspiral and detach from the rest of the $N$-body system.
Such captures are well known and studied in the single-single case \citep[e.g.][]{Hansen:1972il, 1975MNRAS.172P..15F, 1977ApJ...213..183P, 1985AcA....35..401G, 1985AcA....35..119G, 1986AcA....36..181G, 1986ApJ...310..176L, 1993ApJ...418..147L}.  Their outcome could have interesting observational consequences,
from the formation of transients \citep[e.g.][]{2016ApJ...823..113P}, to compact mass transferring binaries \citep[e.g.][]{1975MNRAS.172P..15F, 1975ApJ...199L.143C}.
Interestingly, recent studies indicate that such captures form at a
higher rate during few-body interactions, compared to single-single interactions.
For example, it was recently shown by \cite{2014ApJ...784...71S}, that the rate of eccentric binary black hole (BBH) mergers forming through captures mediated by gravitational wave emission 
likely is dominated by three-body interactions, and not single-single interactions.
Similar eccentric mergers can also form through tidal captures in three-body interactions, as shown by \cite{2008MNRAS.384..376G, 2010MNRAS.402..105G, 2017ApJ...846...36S}.
The point here is that the majority of such eccentric capture mergers are likely to form in dense stellar systems, compared to say the field,
and any observation of such eccentric sources will therefore be an indirect probe of the dynamical channel for BBH and other mergers and the importance of dense stellar environments.
Despite their possible importance, energy loss terms are often not included in the $N$-body equations-of-motion (EOM) \citep[e.g.][]{Fregeau2004}.
For this reason how energy losses during strong few-body encounters, including GWs and tides, affect not only the host cluster dynamics, but also
the range of relevant observables, is not yet well understood.  

Energy dissipation from GW emission is not difficult to include in $N$-body codes thanks to the development of the post-Newtonian (PN) formalism \citep[e.g.][]{2014LRR....17....2B},
and aspects of such corrections have therefore been studied. For example, using full $N$-body simulations \cite{2006MNRAS.371L..45K} showed
how large BHs can form in a GW capture run-away.  Similarly, \cite{2014ApJ...784...71S, 2017ApJ...840L..14S, 2017arXiv171107452S, 2018ApJ...853..140S}
performed isolated three-body scatterings which lead them to conclude that the GW captures forming during the interactions are likely to dominate the rate
of eccentric BBH mergers forming in globular clusters (GCs) observable by the `Laser Interferometer Gravitational-Wave Observatory' (LIGO).
A monte-carlo (MC) approach for studying the evolution of GCs including scatterings up to binary-binary interactions with PN terms was recently
presented by \cite{2017arXiv171204937R}, along with a similar study by \cite{2017arXiv171206186S}, who both confirmed that GW emission in the $N$-body EOM
is crucial for probing the population of eccentric BBH mergers forming in clusters \citep{2017arXiv171107452S}.

Including tides is significantly more difficult than gravitational wave emission.  This is due not only to our limited understanding of stellar structure and the mechanism(s) via which tides are excited and subsequently dissipated,
but also because tidal effects are extremely time consuming to computationally evolve in an $N$-body code.
Some few-body studies have been done using full hydro dynamics \citep[e.g.][]{2010MNRAS.402..105G}, but doing large systematic studies are not yet possible
due to computational limits. Other methods for studying the effects from dynamical tides include the impulsive approximation, where
tidal energy and angular momentum losses are included by simply correcting the velocity vectors at pericenter `by hand' every time two of the $N$ objects
pass very close to each other \citep[e.g.][]{2006MNRAS.372..467B}. A similar approach was also used by \cite{Mardling:2001dl}, and does indeed work.  But,
making such discontinuous corrections to the $N$-body system often lead to poor performance and complicated decision making.
Finally, other approaches include only solving for the evolution of a subset of the tidal modes, which can be done in both linear tidal theory \citep[][]{Mardling:1995hx} using the
Press and Teukolsky (PT) approach \citep{1977ApJ...213..183P}, and non-linearly using the so-called affine model
\citep[e.g.][]{1985MNRAS.212...23C, Luminet:1986cha}; a model we will apply later in this paper. However, such prescriptions are still too
computationally expensive for say parameter space studies and derivations of tidal capture cross sections \citep{2017ApJ...846...36S}; other strategies are therefore needed.

In this paper we propose to include energy loss effects in few-body codes through a simple drag force term in the equations-of-motion. Many few-body codes
have already been optimized to include drag forces, e.g., the 2.5 PN term that accounts for energy dissipation through the emission of GWs is no more than
a simple drag force. Our ansatz in this paper is therefore to derive a general drag force that can be used to model any energy loss effects, and again,
with tides as the main motivation. The only input our drag force model requires is an estimate for the amount of orbital energy lost if two of the $N$
objects undergo a near parabolic encounter. This has been calculated in several studies for tides \citep{1977ApJ...213..183P, 1986ApJ...310..176L},
and fitting formulae have also been provided to speed up these calculations \citep[e.g.][]{1985AcA....35..401G, 1993A&A...280..174P}.

As illustrated in this paper, the use of such fitting formulae together with our proposed drag force model allows one to quickly evolve few-body systems with
both energy losses from tides and GW emission. We note here that our model does not give any new insight
into the two-body tidal problem, but it will be able to provide insight into how especially tides affect the evolution of chaotic few-body interactions.
For that reason, our model has the same limitations as the two-body tidal problem, e.g., we are not able to predict what happens after a tidal capture;
do the two objects merge or do they form a stable binary? However, what we are able to accurately probe and resolve the number of tidal
captures forming in chaotic few-body interactions. We illustrate this in a few examples, by performing controlled two-body and three-body experiments with
different tidal implementations, including our proposed drag force model. In the near future we plan to include this model into the
\texttt{MOCCA} (MOnte Carlo Cluster simulAtor) code \citep{Hypki2013,Giersz2013}, which will allow us to perform systematic studies of how
tidal energy losses in chaotic interactions could affect observables and feedback in to the underlying host cluster dynamics. These are key questions that have to be addressed,
as new searches for transient phenomena will soon be monitoring the sky, including LSST \citep{2009arXiv0912.0201L}, JWST \citep{2006SSRv..123..485G}, and
WFIRST \citep{2013arXiv1305.5422S}.

The paper is organized as follows. In Section \ref{sec:Drag Force Model} we present our drag force model, and describe how to normalize it
for different energy loss mechanisms. We especially discuss how to apply it for describing tidal energy losses, which is
the main motivator for this paper. A short step-by-step description of how to implement the model in an $N$-body code is also given.
In Section \ref{sec:Drag Force Model} we numerically evolve a few two-body and three-body scattering experiments involving tidal and compact objects
with the inclusion of our proposed drag force model. We especially compare our drag force results with other tidal prescriptions, including the impulsive and
affine tidal approximations. Conclusions are given in Section \ref{sec:conclusions}.

\section{Drag Force Model}\label{sec:Drag Force Model}

In this section we describe and derive our proposed tidal drag force model, that in principle can be used to dynamically evolve
chaotic few-body systems with the inclusion of any type of energy loss mechanism; however, our main motivation is how tidal effects impact
the evolution. In short, our approach is to model orbital energy losses by introducing a drag force that acts against the relative motion between any pair of objects
in the few-body system. For deriving the drag force, we assume that the largest energy loss occurs during close pairwise encounters,
and that these encounters can be considered as isolated two-body systems during
the period where most of the energy is lost (see Figure \ref{fig:twobodyill}). This is an excellent assumption, as basically all of the relevant energy loss mechanisms
depend steeply on the relative distance between the objects, implying that most of the energy loss in few-body systems do indeed take place during close pairwise
encounters \citep[e.g.][]{2017ApJ...846...36S}.
The amount of energy that is lost over a single close passage for an isolated two-body system has been studied extensively in the literature, both for
GWs \citep{Peters:1964bc, Hansen:1972il, 1977ApJ...216..610T} and tides \citep{1977ApJ...213..183P, 1985AcA....35..401G, 1985AcA....35..119G}.
Following our assumption that every object pair in the few-body interaction can be treated as an isolated binary when modeling the energy
loss, this allows us to derive the normalization of the drag force, a computation has to be done at each time step.

The functional form of the drag force that controls how energy is lost over a given orbit has to be such that most of the loss
takes place near the pericenter of the two considered objects. This emulates how the energy is actually lost in many mechanisms, and it naturally
makes the assumption of pairwise two-body isolation a good approximation for the purpose of modeling energy losses.
We note that with such a drag force the pairwise energy loss will be close to that found from the impulsive approach, where the individual velocity
vectors are `corrected by hand' at each pericenter passage \citep[e.g.][]{2006MNRAS.372..467B}. However, the loss of energy due to the drag force
will happen continuously over the orbit, which makes our approach both more realistic and easy to implement
in modern few-body codes. In fact, besides the implementation of the drag force into the few-body code described in \cite{2017ApJ...846...36S} that we will
use in this paper, we have already successfully implemented it in the regularized code used in \citep[e.g.][]{2016ApJ...831...61T}

In the sections below we illustrate how a drag force with the properties described above can be constructed. We also discuss its limitations,
and what can be improved.

\subsection{Drag Force Functional Form}\label{sec:Drag Force Functional Form}

\begin{figure}
\centering
\includegraphics[width=\columnwidth]{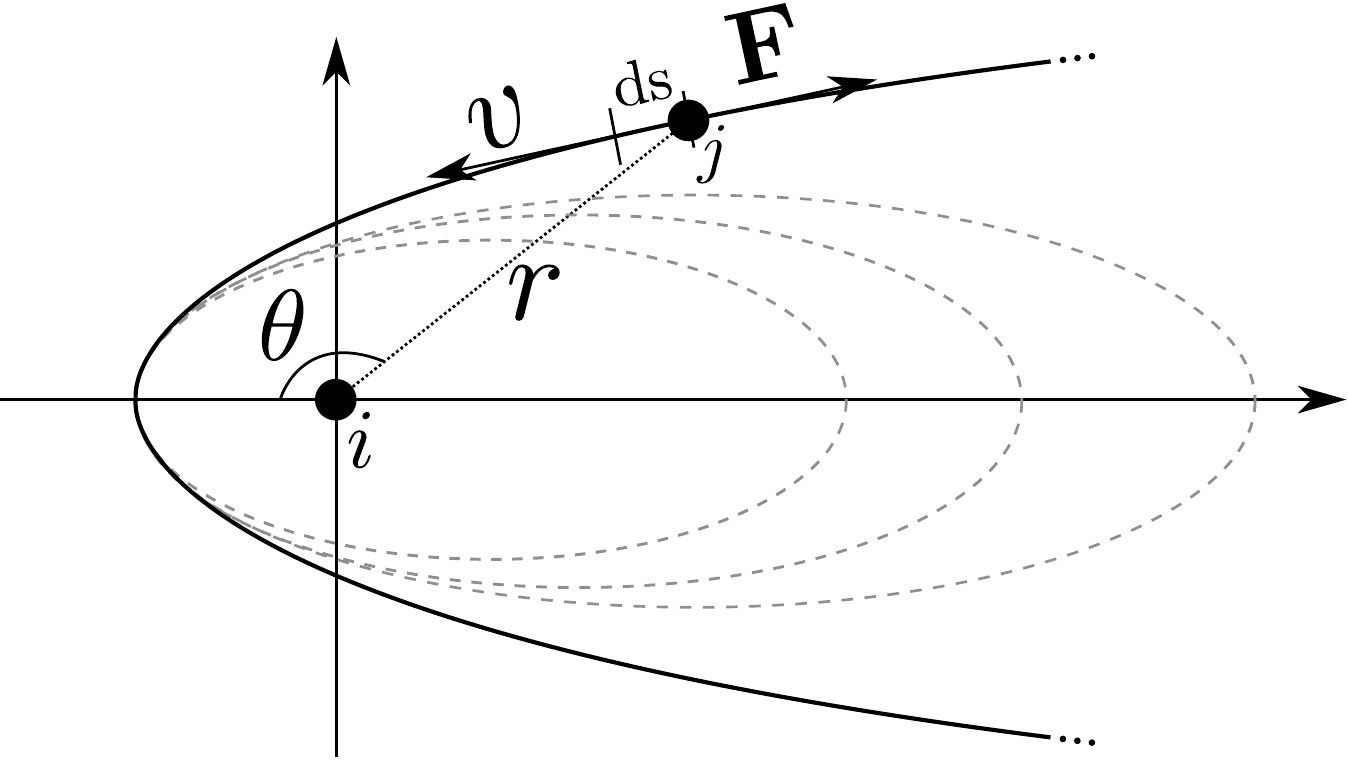}
\caption{Illustration of an eccentric two-body encounter that forms during a chaotic few-body interaction, as described in Section \ref{sec:Drag Force Model}.
If the orbital energy is conserved during the encounter, the two objects, $i$ and $j$, will to leading order pass each other on a near parabolic
Kepler orbit (solid black curve). However, several astrophysical mechanisms are known to lead to orbital energy losses, including tidal excitations and the
emission of GWs. If such mechanisms are included in the EOM, then the two objects can also undergo a dissipative inspiral (the first part of an inspiral
is shown with dashed curves), loosely referred to as a capture, after which they either merge or settle to become a tight semi-stable
binary (see Section \ref{sec:Open Problems in Tidal Captures}). In this paper we propose to include energy losses
through a simple drag force, ${\bf F}$, which acts against the relative orbital motion of the two objects. The figure above illustrates how we define the
quantities $\theta$ (true anomaly), $r$ (relative distance), $v$ (relative velocity),  ${\bf F}$ (drag force), and $ds$ (differential line element),
used in Section \ref{sec:Drag Force Model} for deriving the drag force. The evolution of a tidal capture event described using different tidal
prescriptions is shown in Figure \ref{fig:2bodyex}.
}
\label{fig:twobodyill}
\end{figure}

We consider two objects on a Kepler orbit, bound or unbound, with an initial semi-major (SMA) $a$ and eccentricity $e$.
For this system we now consider a drag force with magnitude $F$ that acts against the orbital motion of the two objects,
as further described and illustrated in Figure \ref{fig:twobodyill}. In this picture, the two-body system will loose an amount of orbital energy $dE = F \times ds$ per differential
line element $ds$ integrated along the orbit. Assuming that the change in orbital angular momentum per orbit is negligible, one can in all relevant cases approximate the
total energy loss over one orbit, denoted by $\Delta{E}$, by the following integral (see \cite{1977ApJ...216..610T} for a similar procedure applied to GW energy losses),
\begin{equation}
\Delta{E} \approx \int_{-\theta_{0}}^{+\theta_{0}} F \frac{ds}{dt} \frac{dt}{d\theta}d\theta,
\label{eq:intDE}
\end{equation}
where $dt$ is the differential change in time, $\theta$ is the true anomaly, and $\theta_{0} = \pi$ for a bound orbit and $= \cos^{-1}(-1/e)$ for
an unbound orbit (see Figure \ref{fig:twobodyill}). The term $ds/dt$ is simply the relative velocity between the two interacting objects, denoted by $v$, which can be written as,
\begin{equation}
\frac{ds}{dt} = v = \left[ 2GM \left( \frac{1}{r} - \frac{1}{2a} \right)\right]^{1/2},
\end{equation}
where $M = m_{i} + m_{j}$ is the total mass of the two interacting objects, referred to as $i$ and $j$, and $r$ is their relative distance.
The term $dt/d\theta$ can be derived from Kepler's relation $r = a(1-e^2)/(1+ e\cos\theta)$, from which it follows,
\begin{equation}
\frac{dt}{d\theta} = \frac{[a(1-e^2)]^{3/2}}{(GM)^{1/2}} \frac{1}{(1+e\cos\theta)^{2}}.
\end{equation}
To proceed we now have to chose a functional form for the drag force, ${\bf F}$. As described, the form should both be simple to implement in a few-body code,
possibly similar to the 2.5 PN drag force that has been successfully implemented in many recent few-body codes, while ensuring that most
of the energy loss happens at pericenter. A first proposed form could be a force that is $\propto 1/r^{n}$, where $n$ is some power; however,
in this case one finds that the integral in Equation \eqref{eq:intDE} does not have an analytical solution for any $n$, including $n=0$. This `problem' relates to the fact that
the circumference of an ellipse cannot be written out in a closed form. This is why `elliptical' integrals always have to be solved numerically.
However, if we instead choose a force that is $\propto v/r^{n}$, then the integral in Equation \eqref{eq:intDE} can be written out in closed form for any $n$, which
allows us to analytically estimate the drag force normalization, or coefficient. This leads to a very fast derivation of the drag force per time step.
In this paper we therefore choose to work with the following drag force,
\begin{equation}
{\bf F} = - {\mathscr{E} \frac{v}{r^{n}}} \times \frac{{\bf v}}{v},
\label{eq:DF_eq}
\end{equation}
where $\mathscr{E}$ is a normalization factor that to leading order depends on the orbital parameters for the two-body system and the considered energy loss
mechanism. Although this choice of drag force could seem arbitrary, we note that the 2.5 PN drag force is exactly of this type with $n=4$.
Therefore, a code that is optimized to run with PN terms, should have no problem in evolving the system with our proposed drag force.
Below we illustrate how to estimate the drag force coefficient $\mathscr{E}$.

\subsection{Drag Force Normalization Coefficient}

The coefficient $\mathscr{E}$ of the drag force introduced in the above Equation \eqref{eq:DF_eq}, can be estimated using Equation \eqref{eq:intDE} assuming
that $\Delta{E}$ is known a priori for the considered two-body system. After some algebraic manipulations we find from solving equation \eqref{eq:intDE} with
our proposed drag force from Equation \eqref{eq:DF_eq} that,
\begin{equation}
\mathscr{E} = \Delta{E} \times \frac{1}{2} \frac{\left[ a(1-e^2) \right]^{n-1/2}}{\left( GM \right)^{1/2} \mathscr{I}(e,n)},
\label{eq:EPS_gennorm}
\end{equation}
where $\mathscr{I}(e,n)$ is the solution to the following integral
\begin{equation}
\mathscr{I}(e,n) = \int_{-\theta_{0}}^{+\theta_{0}}\frac{(1+e\cos\theta) - (1-e^2)/2}{(1+e\cos\theta)^{2-n}} d\theta.
\label{eq:I_int_en}
\end{equation}
This factor $\mathscr{I}(e,n)$ can be written out in closed form for any value of $n$. In this paper we will study
the performance for two different values of $n$, namely for $n=4$ and $n=10$. For these two cases $\mathscr{I}$ evaluates to,
\begin{equation}
\mathscr{I}(e,n=4,10) = 
\begin{cases}
\frac{\pi}{2}\left(2 + 7e^2 + e^4 \right), & \text{for}\ n=4  \\[2ex]
\frac{\pi}{128} (128 + 2944e^2 + 10528e^4 \\
+ 8960e^6 + 1715e^8 + 35e^{10}), & \text{for}\ n=10,
\end{cases}
\label{eq:I_n4n10}
\end{equation}
where we have assumed that  $\theta_{0} = \pi$. In the unbound case for which $\theta_{0} = \cos^{-1}(-1/e)$, the operation $\cos^{-1}$ has to be performed
at each time step, which is computationally heavy. However, for all practical purposes, there is no problem in just using the value of $\mathscr{I}$ assuming
that $\theta_{0} = \pi$, i.e., use relations similar to the one shown in Equation \eqref{eq:I_n4n10}, even when the objects are not bound initially. The reason is simply that
if the considered energy loss mechanism turns out to be significant, then the two objects in question will immediately get bound to each.
For all the numerical simulations shown in this paper, we will therefore assume the bound limit when calculating the factor $\mathscr{I}$; that is, we we will
use the relations from Equation \eqref{eq:I_n4n10}. Below we will describe how $\Delta{E}$ can be estimated for tides, and discuss open problems related to our adopted methodology.

\subsection{Orbital Energy Losses}

Our introduced drag force given by Equation \eqref{eq:DF_eq}, is only a convenient kernel that leads to a continuous loss of orbital energy.  It therefore needs to be normalized at each time step. As seen from Equation \eqref{eq:DF_eq} and \eqref{eq:EPS_gennorm}, the normalization happens
through the parameter $\mathscr{E}$, which depends on the expected two-body energy loss over one orbit, $\Delta{E}$.
From Equation \eqref{eq:EPS_gennorm}, it is clear that the evaluation of $\Delta{E}$ is the only computation that potentially could take up a significant amount of
CPU time; however, in this section, we point out that in most cases very simple analytical formulae or functional fits exist which can by-pass this expense.
One example is the orbital energy loss through GWs, which was shown by \cite{Hansen:1972il} to simply
scale with the two-body pericenter distance, $r_{\rm p}$, as $\propto r_{\rm p}^{-7/2}$. For tides, which is our main motivator for this
paper, the form is more complicated and will therefore be described in greater detail below. We will also
comment on potential problems and open questions when describing strong tidal interactions over subsequent passages with this approach.

\subsubsection{Tidal Energy Loss}\label{Sec:Tidal Energy Loss}

Consider a two-body encounter between a tidal object `t', and a perturber `p'.
In the linear limit to quadruple order ($l = 2$) the energy loss integrated over an unperturbed parabolic orbit can be written as \citep{1977ApJ...213..183P},
\begin{equation}
\Delta{E} \approx \left( \frac{GM_{\rm t}^2}{R_{\rm t}} \right) \left( \frac{M_{\rm p}}{M_{\rm t}} \right)^{2} \left( \frac{R_{\rm t}}{r_{\rm p}} \right)^{6} \times T(\eta),
\label{eq:dE_tides}
\end{equation}
where $r_{\rm p}$ is the initial orbital pericenter distance $ = a(1-e)$, $T(\eta)$ is a function with no analytic solution that depends on the internal
structure of the tidal object in question, and $\eta$ is a parameter that is defined by,
\begin{equation}
\eta = \left( \frac{M_{\rm t}}{M_{\rm t} + M_{\rm p}}\right)^{1/2} \left( \frac{r_{\rm p}}{R_{\rm t}}\right)^{3/2}. 
\end{equation}
The above Equation \eqref{eq:dE_tides} accounts for the energy loss when only one of the objects is a tidal object. If both of the objects
are tidal objects, then the total energy loss is simply the sum of the pairwise energy losses \citep[see, e.g.,][]{1993A&A...280..174P}.
Now, the time consuming part in calculating the tidal energy loss $\Delta{E}$ from above, is to calculate the value of $T(\eta)$.
To avoid this, fitting functions have been made to approximate $T(\eta)$ for different stellar
polytropes \citep[e.g.][]{1985AcA....35..401G, 1993A&A...280..174P}. By the use of such fitting functions,
the tidal energy loss over one orbit can therefore quickly be estimated
for any combination of objects by the use of Equation \eqref{eq:dE_tides}. For the examples studied in this paper, we use the fitting
formulae presented in \cite{1993A&A...280..174P}, which allows us to quickly estimate $\Delta{E}$ from $\mathscr{E}$, and thereby the drag force at each time step.
This strategy is on average expected to give reasonable results; however, there are several well known problems related to this description of two-body tidal captures.
A few of these problems will be discussed below.

\subsubsection{Open Problems in Describing Tidal Captures}\label{sec:Open Problems in Tidal Captures}

For describing tidal interactions with our drag force model, one needs to know the exact tidal energy loss over each orbit for each object pair.
However, even the evolution of a simple isolated two-body tidal capture is currently poorly understood \citep[see, e.g.,][]{2017MNRAS.467.4180S}. Aspects of this problem
relate especially to how fast, and in what way(s), the tidal modes that have been excited during the pericenter passage damp. If the modes damp
quickly, and the deposited energy is radiated away shortly thereafter, then the energy loss at each pericenter passage will be independent of previous passages.
However, if say the tidal modes do not effectively damp between pericenter passages, then the energy loss over a given orbit depends on all previous passages.
The reason is that if the tidal modes are not damped, they will couple to the orbital motion through a quadrupole interaction
term induced by the tidal bulge \citep[e.g.][]{1992ApJ...385..604K, 2017ApJ...846...36S}.  This implies
that the orbital energy stored in tidal oscillations can be put back into the orbit at later times. This can give rise to highly chaotic
orbital motions \citep{1992ApJ...385..604K, Mardling:1995hx, Mardling:1995it, 2017ApJ...846...36S}.
If the tidal modes damp quickly, but the energy is not radiated away on short time scales, the tidal object in question will likely undergo a momentary expansion,
which might lead to a runaway tidal capture followed by a merger. Therefore, the evolution of the capture depends strongly on how the tidal modes damp,
and how fast the energy is dissipated.

Another issue relates to how the angular momentum is transferred from the orbit to the stellar object(s), and how it is later dissipated.
For example, if a tight binary forms as a result of a capture with an initial pericenter distance $r_{\rm c}$, then the binary will end up on a circular orbit with a
final SMA $ = 2r_{\rm c}$ if only little angular momentum is dissipated; however, if the angular momentum instead is effectively dissipated, then the final SMA will
be smaller, which might lead to unstable mass transfer followed by a merger.

The two-body tidal captures we consider in this work that are formed in resonating few-body systems, are likely to be further affected by secular
effects. The reason is that if a tidal capture binary forms during a resonating few-body interaction, then
it will in principle undergo its evolution with at least one bound companion \citep[e.g.][]{2017ApJ...846...36S}, which can lead to secular evolution of the binary
through the Kozai-Lidov mechanism \citep[e.g.][]{1962AJ.....67..591K, 2016ARA&A..54..441N}.
This implies that even if the two objects in question undergo a tidal capture that first results in a stable binary, they might be driven to
merger shortly after by the remaining bound tertiary object(s). This motivates looking into the tidal capture problem in the presence of bound tertiary objects. 

Finally, a technical issue, which can be solved but has not been addressed in detail yet, is the amount of energy dissipated over one orbit when the
two objects are bound, i.e., not a parabolic orbit as assumed in \cite{1977ApJ...213..183P}. In the bound case one expects that the efficiency of the energy dissipation should decrease
with decreasing eccentricity $e$, asymptoting to zero for circular orbits $e = 0$. An attempt to modify the PT model to also work for bound orbits was presented in \cite{Mardling:2001dl},
which showed that a simple redefinition of $\eta$ can be used. Another paper by \cite{1993A&A...280..174P}, suggests to simply multiply the PT value by $e$.
So, although the problem has not been studied in detail yet, the work discussing this so far indicates that including the effect into our fitting approach is not difficult.

Despite these issues and complications, it was recently illustrated by \cite{2017ApJ...846...36S}, that for at least resonating three-body interactions, the main
effect from including tides is indeed the formation of tidal captures that form during the interactions. This was shown using the affine model with no mode damping,
which allows for strong tidal and orbital couplings.  In principle, this coupling could lead to resonating and chaotic behaviors that prevent the formation of captures;
however, that was not clearly observed in the simulations. We therefore expect that our simple drag force model will predict at least the right number of binary captures. If these binaries will subsequently remain stable or merge is not currently clear, as described above. 

In the next section, we describe how to implement our proposed drag force model into an $N$-body code.

\subsection{$N$-body Implementation}

Consider an object pair denoted by $i,j$ in an $N$-body system. In the following we describe step by step how to derive the drag force from Equation \eqref{eq:DF_eq},
exerted by object $j$ on object $i$. This has to be done at each time step. The procedure is as follows:
\begin{itemize}
\item Choose a value for the drag force parameter $n$, which determines the steepness of the drag force near
pericenter, and from that calculate the analytical expression for $\mathscr{I}$ given by Equation \eqref{eq:I_int_en}.
\item From the position and velocity vectors of the two objects, calculate their (oscullating) SMA $a$, and
eccentricity $e$, assuming they are on an isolated Kepler orbit.
\item From the derived $a,e$ estimate how much energy should in theory be lost over such an (isolated two-body)
orbit, $\Delta{E}$.  For tides, this can be done following the prescriptions in Section \ref{Sec:Tidal Energy Loss}. Calculate also the value of $\mathscr{I}$ given $e$.
\item From $a,e$, $\Delta{E}$, and $\mathscr{I}$, estimate the drag force coefficient $\mathscr{E}$ by the use of Equation \eqref{eq:EPS_gennorm}.
\item Using the value for $\mathscr{E}$ write out the drag force vector between the two objects, here refered
to as ${\bf F}_{\rm ij}$, via the use of Equation \eqref{eq:DF_eq}.
\item The final acceleration of object $i$ exerted by object $j$, due to whatever energy loss mechanism that is considered, is now simply
given by ${\bf a}_{\rm ij} = {\bf F}_{\rm ij}/m_{\rm i}$. 
\end{itemize}

This procedure is repeated for all objects $j$, and the total drag force acceleration of object $i$ is then simply the corresponding vector sum.
This total acceleration is then added to the remaining acceleration terms in the $N$-body code, which of course includes the Newtonian acceleration,
and possibly also the PN terms. We note here that all of the quantities that are needed for calculating the drag force acceleration, must also be derived for the $2.5$ PN term. The drag force comes therefore at nearly no additional computational cost if the $2.5$ PN term is already included. 

In the sections below, we study the performance of the drag force model in the case of tidal energy losses, by running a few examples, using the
implementation procedure described above. For this we use the few-body code presented in \cite{2017ApJ...846...36S}. Assumptions, limitations,
and general results are described in the following.

\section{Comparisons and Examples}

In the following sections we study and describe the role of tides in two-body and three-body interactions.
We especially compare results derived using different tidal prescriptions, including the affine tides model \citep{Luminet:1986cha, 2017ApJ...846...36S}, our
drag force model with $n=4,10$ (see Section \ref{sec:Drag Force Functional Form}), and a model from which an analytic solution can be written out in closed form.

\subsection{Two-body Interactions with Tides}\label{sec:Two-body Interaction with Tides}

\begin{figure}
\centering
\includegraphics[width=\columnwidth]{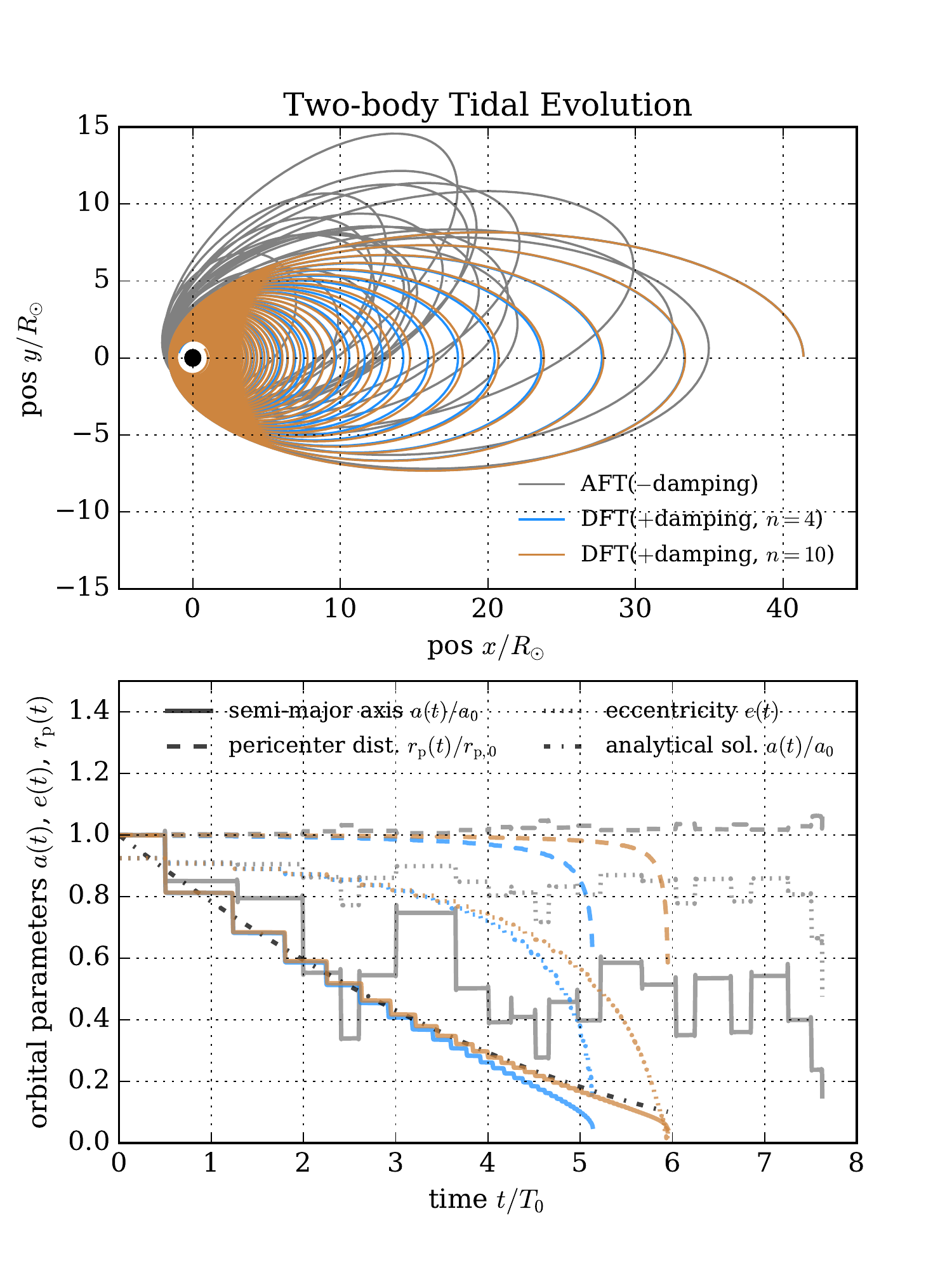}
\caption{Study of the tidal evolution of an eccentric binary consisting of a
solar type star, modeled as a polytrope with index $3$, mass $1M_{\odot}$, and radius $1R_{\odot}$,
and a compact object with mass $1.4M_{\odot}$. The initial SMA is $a_{0} = 0.1$ AU, and pericenter distance $\approx 1.5R_{\odot}$.
Both figures show results from three simulations with the same initial conditions (ICs), but different tidal prescriptions.
As explained in Section \ref{sec:Two-body Interaction with Tides}, the {\it grey} curves show results using the affine model with no damping,
whereas the {\it blue} and the {\it orange} curves show results from our proposed drag force model with $n=4$ and $n=10$, respectively.
These three tidal prescriptions are labeled in the figures by `AFT($-$damping)', `DFT($+$damping, $n=4$)', and `DFT($+$damping, $n=10$)'.
{\it Top plot}: Evolution shown in the orbital plane, with one of the two objects fixed at $0,0$. As seen, all three models lead to an overall
decrease in the SMA due to the exchange of orbital energy into tidal excitations. The affine model shows indications of chaotic motion
and orbital precession; these chaotic effects are due to the tidal modes coupling to the orbit.
{\it Bottom plot}: The corresponding evolution of the orbital parameters as a function of time $t$, in units of the initial orbital time $T_{0}$. The parameters
shown are the SMA $a(t)$ (solid line), the pericenter distance $r_{\rm p}(t)$ (dashed line), and the eccentricity $e(t)$ (dotted line). The
{\it dash-dotted} line shows the analytical solution given by Equation \eqref{eq:at_analy}. As seen, both tidal drag force prescriptions ($n=4,10$)
lead to a smooth decay, which follows the analytic solution very well. The affine model leads instead to chaotic behavior after a few
pericenter passages. On average, or if damping is efficient, the affine model is expected to accurately follow the drag force decay.
Further descriptions are given in Section \ref{sec:Simulations and Results}.
}
\label{fig:2bodyex}
\end{figure}

In this section, we perform a controlled numerical experiment to quantify how the orbital parameters of a fully isolated two-body system, consisting of a 
tidal object and a perturber, evolve as a function of time when orbital energy losses through tides are included.
As described below, we do this using a few different tidal prescriptions, including our proposed tidal drag force model described in Section \ref{sec:Drag Force Functional Form}.
We note that several more detailed studies of the two-body tidal problem have been
performed \citep[e.g.][]{1992ApJ...385..604K, Mardling:1995hx, 2018MNRAS.476..482V}; however, our study is
mainly performed to validate our derived drag force model under the imposed assumptions, while also quantifying how our prescription both depends on the free parameter $n$, and compares to other models.

\subsubsection{Initial Conditions}

We consider the interaction between a tidal object (solar type star modeled as a polytrope with index $3$, mass $1M_{\odot}$, and radius $1R_{\odot}$) and a
compact object ($1.4M_{\odot}$ point mass, which could represent a neutron star; NS). The two objects are initially located at apocenter on an orbit with initial SMA $a_{0} = 0.1$ AU,
and pericenter distance $\approx 1.5R_{\odot}$, which corresponds to an eccentricity of $\approx 0.93$. This could represent a temporary binary formed during a resonating
few-body interaction, often referred to as an intermediate state (IMS) binary \citep[e.g.][]{2014ApJ...784...71S}. In the following we study the evolution of this binary
when tides are taken into account.

\subsubsection{Simulations and Results}\label{sec:Simulations and Results}

The orbital evolution of the two-body system described above is shown in Figure \ref{fig:2bodyex}, and described in the following.
The `grey' line (labeled 'AFT($-$damping)', where 'AFT' is short for `affine tides'), shows the system evolved using the affine model
with no mode damping included. As described in \cite{Luminet:1986cha}, in the affine model the tidal object in question is modeled as a triaxial polytrope, which is coupled to the
orbital motion of the few-body system through its center-of-mass (COM) and its quadrupole force induced by the
tidal excitation \citep[see also][]{1995MNRAS.275..498D}. This allows for tidally stored energy to be
pumped back into the orbit if the tidal object is excited before the encounter. The `blue' (`orange') line shows the system evolved using the
drag force model described in Section \ref{sec:Drag Force Functional Form} with $n=4$ ($n=10$), and $\Delta{E}$ derived using the PT model
assuming the parabolic limit. The `dash-dotted' line shows the analytic solution to the time evolution of the SMA assuming the system
loses a constant amount of energy per pericenter passage.  This change in energy is set equal to the energy lost during the first passage, referred to here as $\Delta{E}_1$.
To facilitate easy comparisons, we use the parabolic PT model to calculate $\Delta{E}_1$. The analytic solution was found by assuming the orbit averaged limit, from which
the change in orbital energy $dE(t)$ per unit time $dt$ can be written as \citep[e.g.][]{2018ApJ...853..140S},
\begin{equation}
\frac{dE}{dt} \approx \frac{\sqrt{2}}{\pi} \frac{\Delta{E}_1}{GM}\mu^{-3/2}E(t)^{-3/2},
\end{equation}
where $\mu$ is the reduced mass. This equation can be solved for $E(t) = GM\mu/(2a(t))$, which can then be rewritten in terms of
the SMA $a(t)$ of the system as,
\begin{equation}
a(t) \approx \left( \sqrt{a_{0}} - \gamma t \right)^{2},
\label{eq:at_analy}
\end{equation}
where
\begin{equation}
\gamma = \frac{1}{2\pi}\sqrt{M\mu}\frac{\Delta{E}_1}{M}\mu^{-3/2}.
\end{equation}
This solution is expected to describe the orbital evolution relatively accurately in the limit where the energy transferred between the orbit and the tidal object
during each pericenter passage are uncorrelated, i.e., when damping and dissipation are efficient. However, it will break down when the system starts to circularize,
which from Equation \eqref{eq:at_analy} will happen after a time $\approx \sqrt{a_{0}}/\gamma$.
This is also known as the inspiral or life time of an eccentric system with energy losses \citep[e.g.][]{Peters:1964bc}.

By comparing these different numerical solutions, we first notice that the two versions of the drag force model ($n=4,10$) and the analytic solution
from Equation \eqref{eq:at_analy} follow each other accurately during the first stages of the tidal inspiral evolution. However, at late times, just before circularization, the $n=4$
version (blue line) starts to deviate from the other two solutions. The reason is that for lower values of $n$ the energy loss is
smeared out over a larger fraction of the orbit leading to a derived pericenter distance that decreases notably over one orbit.
This leads to an increase in the energy loss over the orbit, which in turn makes the system
evolve faster near the point of circularization.  This explains why the $n=4$ version leads to a slightly faster merger. For higher values of $n$,
such as the $n=10$ version considered here, the loss of energy becomes increasingly more `impulsive', in the sense that it is removed over a smaller fraction of the orbit is more concentrated 
near pericenter. For $n \rightarrow \infty$ the drag force model therefore approaches the impulsive limit. The evolution in that limit is close to
the analytic solution given by Equation \eqref{eq:at_analy}.  This explains why the $n=10$ version
follows the analytic prediction better than the $n=4$ version.

We now move on to the affine model. As seen in the figure, the strong coupling between the excited modes and the orbit 
lead to highly chaotic motions. After a few passages, where the affine model actually traces the other models recently well,
the evolution enters a semi-chaotic phase.  This leads the system to be far from undergoing a merger at the point when the drag force model
predicts a merger ($t/T_{\rm 0} \approx 5-6$). This is not a generic feature of the affine model, as it also sometimes happens that a merger occurs before
the point found from the drag force model. In general, large changes in the outcome of the affine model are found by only marginally changing the initial conditions.  This is mainly due 
to the `orbital time' of the $l=2$ mode of the star being much shorter than the Keplerian orbital time of
the two interacting objects (for a systematic study on this see, e.g., \cite{2018MNRAS.476..482V}). However, despite this semi-chaotic behavior, one still 
expects the average decrease in SMA to follow the evolution found from assuming instantaneous damping and dissipation (i.e., close to our tidal drag force model),
where the dispersion around this average should scale with the number of pericenter passages \citep[e.g.][]{1992ApJ...385..604K}.

If the chaotic evolution induced by the mode-orbit coupling turns out to play on important role in the formation of tidal capture binaries, one can 
include this effect by simply adding a harmonic term.  This additional term should be added to our presented drag force term, with a period close to that of the $l=2$ mode.

Below, we study the formation of two-body capture binaries assembled during three-body interactions.

\subsection{Three-body Interactions with Tides}\label{sec:Three-body Interaction with Tides}

The first systematic study on the chaotic three-body problem with the inclusion of tides was performed by \cite{2017ApJ...846...36S} using the affine model
for evolving the tides. As described in \cite{2017ApJ...846...36S}, the main effect from including tides in the chaotic three-body problem is the formation of tidal capture
binaries that form during the interaction. As described in Section \ref{sec:Open Problems in Tidal Captures}, whether or not the tidally formed binaries promptly merge
or settle onto a semi-stable orbit after reaching the point of circularization is still an open question. Independent of the exact outcome, however, tidal capture binaries do clearly form in chaotic interactions.
The question is, how often do such tidal captures form, and are they dynamically or observationally important? As stated in the introduction,
no clear studies have to date been performed to properly address these questions.  This is mostly because tides are computationally heavy to implement in an $N$-body simulation. Our hope is that our drag
force model, which is both easy to implement and very fast to evolve, can be used to gain further insight into this. 

Below, we consider the evolution of a chaotic three-body interaction, and study how the inclusion of tides affects the outcome.

\begin{figure}
\centering
\includegraphics[width=\columnwidth]{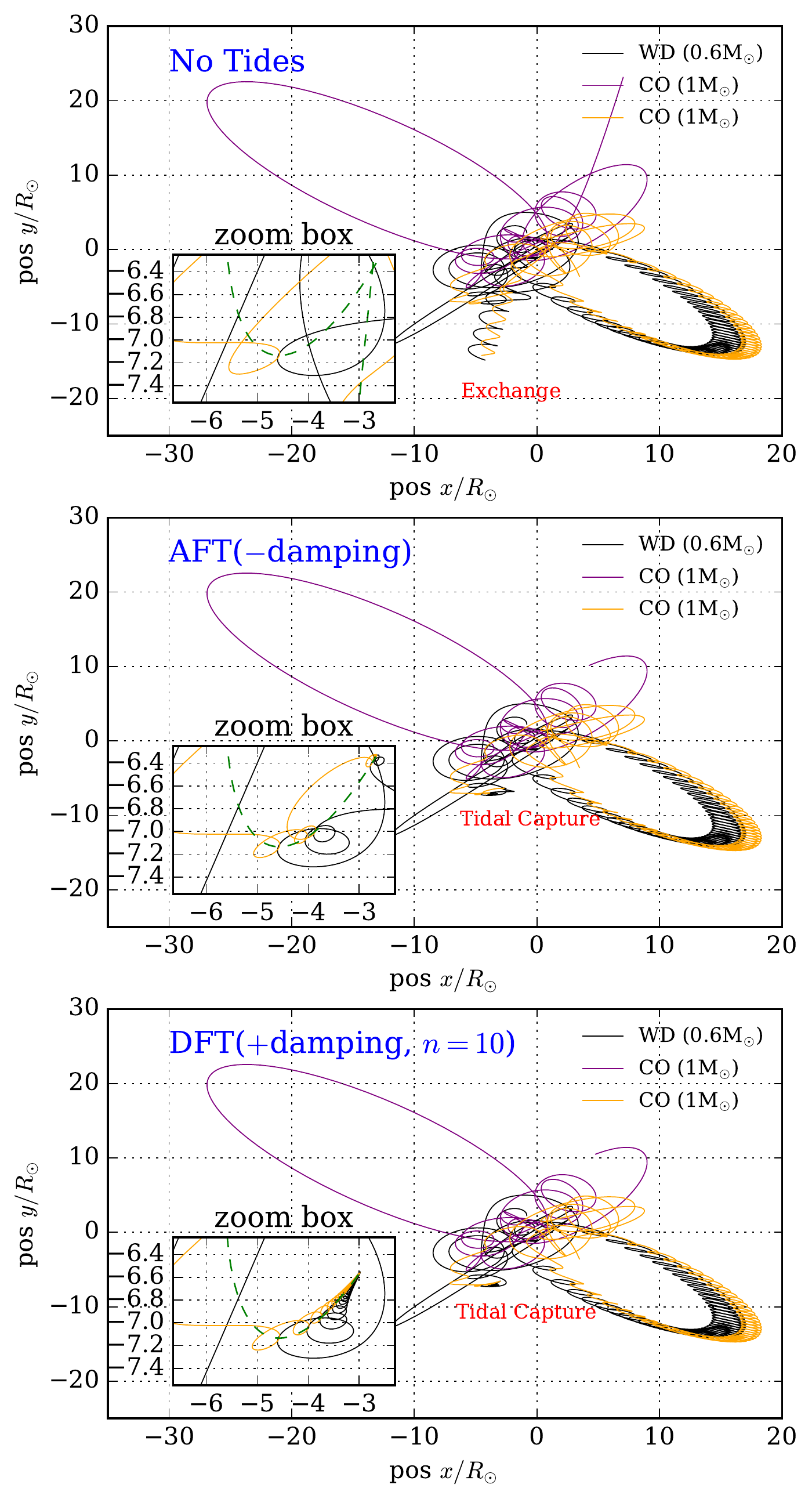}
\caption{Results from binary-single interactions evolved with and without tides included in the EOM.
All three plots have the same ICs, and show the orbital evolution of a binary.  The binary consists of a
white dwarf (black line) modeled as a polytrope with index $3/2$, mass $0.6M_{\odot}$, and radius $0.0136R_{\odot}$,
and a compact object (purple line) with mass $1M_{\odot}$, interacting with an incoming similar CO (yellow line). The initial binary eccentricity is $=0$, and the SMA is $ = 0.01$ AU.
In the {\it top plot} (labeled `No Tides'), tides are not included in the EOM, and the WD is therefore
included as a simple solid sphere. In this case the interaction ends with a classical exchange.
In the {\it middle plot} (labeled `AFT($-$damping)'), tides on the WD are included using the affine model without damping.
The interaction in this case ends instead with a tidal capture between the WD and one of the COs (yellow line). The insert box shows a zoom
in on the tidal capture event, where the green dashed line illustrates the COM trajectory of the interacting WD-CO pair (the same zoomed in region is shown in
each insert box for the three figures). As seen, the inspiral is not smooth, but semi-chaotic.  The chaotic component of the evolution is due to strong couplings between the tidal modes and the orbit.
In the {\it bottom plot} (labeled `DFT($+$damping, $n=10$)'), tides on the WD are modeled using our drag force model with $n=10$ described
in Section \ref{sec:Drag Force Model}. In this case the interaction not only ends with a capture, but the location and evolution of the capture seem to be almost
identical to the interaction integrated using the affine tides (middle plot). This validates our approach and tidal implementations.
We emphasize that the drag force model is much faster to evolve than say the affine model, and much easier to implement.}
\label{fig:3bodyex}
\end{figure}

\subsubsection{Initial Conditions}

We consider a binary consisting of a white dwarf (`WD', modeled as a polytrope with index $3/2$, mass $0.6M_{\odot}$, and radius $0.0136R_{\odot}$)
and a compact object (`CO', with mass $1M_{\odot}$) with an initial SMA of $0.01$ AU, interacting with an incoming identical CO.
We perform three scatterings with exactly the same ICs. In the first interaction we do not include any tidal effects, and treat
the WD as a solid sphere. In the second interaction, we include tides on the WD using the affine model without damping.
In the third interaction, we include tides using our drag force model with $n=10$, assuming $\Delta{E}$ from the parabolic PT model using
the fitting formulae presented in \cite{1993A&A...280..174P}. We describe our results below.

\subsubsection{Simulations and Results}

The orbital evolution of the three binary-single interactions we performed are shown in Figure \ref{fig:3bodyex}.  The top plot shows the results with no tides (labeled `No Tides'),
the middle plot shows the results assuming affine tides with no damping (labeled `AFT($-$damping)'), and the bottom plot shows the results with tides modeled using our drag
force model with $n=10$ (labeled `DFT($+$damping, $n=10$)'). As seen, when tides are not included, the interaction concludes with a classical exchange outcome
in which one of the COs is ejected from the system, leaving behind a WD-CO binary.  As seen in the middle and bottom plots, when tides are included
the interaction instead ends with the WD undergoing a tidal capture event with one of the COs as a result of a very close encounter between the two bodies during the
interaction. Upon comparing the results from the two tidal models, we see that the tidal capture event happens at almost exactly the same point during the interaction.  This 
serves as a good validation of both tidal implementations. However, there are small differences.  These differences are most clearly seen upon considering the zoom boxes shown in
each plot. For example, in the affine model, the inspiral is clearly chaotic.  This chaotic behavior arises from the tidal modes coupling to the orbit.
This is contrast to the evolution from the tidal drag force model, which by construction leads to a smooth inspiral.
In all three zoom boxes are shown the COM motion of the WD-CO binary via a `green dashed' line. As should be the case, both of the tidal models and the
no tides model all give rise to the same COM motion of the considered WD-CO binary.

\section{Conclusions}\label{sec:conclusions}

Accurate modeling of the dynamical assembly of BBH mergers, as well as stellar mergers and disruptions,
requires $N$-body codes that include orbital energy losses in the EOM, such as occur due to tidal excitations and dissipation \citep[e.g.][]{2017ApJ...846...36S}, as
well as due to GW emission \citep[e.g.][]{2006ApJ...640..156G, 2014ApJ...784...71S, 2017arXiv171107452S, 2017arXiv171206186S, 2017arXiv171204937R}.
For example, as recently shown by \cite{2017arXiv171107452S}, if GW emission is included in the EOM then the estimated
rate of eccentric BBH mergers forming in globular clusters is $\sim 100$ times higher than one finds using a standard Newtonian code; a correction that can
play a key role in how to observationally distinguish between different BBH merger channels.
Such PN corrections are relatively easy to implement in $N$-body codes via the use of the PN formalism \citep{2014LRR....17....2B}.
This is in contrast to energy losses from dynamical tides, which are much more challenging to include due to both theoretical and numerical limitations, as described
in Section \ref{sec:Open Problems in Tidal Captures}. For that reason, only limited work has been done on how dynamical tides affect
the dynamics and observables of dense stellar systems.
Some of the work that has been done includes full hydro simulations of three-body encounters \citep{2010MNRAS.402..105G}, $N$-body simulations with `impulsive' tidal
corrections \citep{Mardling:2001dl, 2006MNRAS.372..467B}, and scatterings using linear \citep[e.g.][]{Mardling:1995hx} as well as non-linear \citep[e.g.][]{2017ApJ...846...36S} tidal
models.  However, all of these prescriptions are either too computationally expensive or too decision heavy for systematic parameter space studies. Something more simple and fast is needed, at least for the
few-body codes that currently are used in Monte Carlo studies of dense stellar systems \citep[e.g.][]{2013MNRAS.429.1221H}. 

To overcome the problems related to the inclusion of especially tidal energy losses, we have 
described in this paper a method for including these effects in the $N$-body EOM through a simple drag force prescription, given that the amount of energy loss per close encounter is
known a priori for every object pair in the system. We point out that pair-wise energy losses have been derived
in the literature; for example, for the tidal examples
presented in this paper (See Section \ref{sec:Two-body Interaction with Tides} and \ref{sec:Three-body Interaction with Tides})
we used the two-body tidal fitting functions given by \cite{1993A&A...280..174P}.  This allows us to compute the magnitude and direction
of our proposed drag force almost instantaneously at each time step. This in turn leads to an increase in computational speed by several orders of magnitude compared to previous methods,
such as the affine model \citep[e.g.][]{2017ApJ...846...36S}, and a much simpler implementation than say the impulsive method used in \citep{2006MNRAS.372..467B}.

Our drag force model is particularly well-suited to accurately probe the number of tidal captures forming in chaotic few-body interactions,
including binary-single and binary-binary interactions. However, our model is not suitable for modeling secular systems such as Kozai triples; in this case
other methods should be used \citep[e.g.][]{2007ApJ...669.1298F, 2009ApJ...699L..17P, 2016ARA&A..54..441N}. In upcoming work we
plan to use our new drag force formulation to estimate cross sections for tidal captures forming in chaotic few-body systems, involving different
compact and tidal objects. We are further in the process of including
the model in the \texttt{MOCCA} code \citep{Hypki2013,Giersz2013}, which soon will allow us to perform systematic studies on the
dynamical and observational consequences of tidal and GW energy losses in dense stellar systems.  These studies are essential in preparing for data coming from future and current transient surveys.

\section*{Acknowledgements}
It is a pleasure to thank M. MacLeod, E. Ramirez-Ruiz, M. Giersz, M. Spera, N. Stone,
A. Generozov, and A. Hamers for helpful discussions.
The authors further thank the Center for Computationally
Astrophysics (CCA), where this work was initiated.
J. S. acknowledges support from the Lyman Spitzer Fellowship.
N. W. C. L. acknowledges support from a Kalbfleisch Fellowship
at the American Museum of Natural History and
the Richard Gilder Graduate School, as well as support from
National Science Foundation Award AST 11-09395.
A. A. T. acknowledges support from JSPS KAKENHI Grant Number 17F17764.


\begin{thebibliography}{}
\makeatletter
\relax
\def\mn@urlcharsother{\let\do\@makeother \do\$\do\&\do\#\do\^\do\_\do\%\do\~}
\def\mn@doi{\begingroup\mn@urlcharsother \@ifnextchar [ {\mn@doi@}
  {\mn@doi@[]}}
\def\mn@doi@[#1]#2{\def\@tempa{#1}\ifx\@tempa\@empty \href
  {http://dx.doi.org/#2} {doi:#2}\else \href {http://dx.doi.org/#2} {#1}\fi
  \endgroup}
\def\mn@eprint#1#2{\mn@eprint@#1:#2::\@nil}
\def\mn@eprint@arXiv#1{\href {http://arxiv.org/abs/#1} {{\tt arXiv:#1}}}
\def\mn@eprint@dblp#1{\href {http://dblp.uni-trier.de/rec/bibtex/#1.xml}
  {dblp:#1}}
\def\mn@eprint@#1:#2:#3:#4\@nil{\def\@tempa {#1}\def\@tempb {#2}\def\@tempc
  {#3}\ifx \@tempc \@empty \let \@tempc \@tempb \let \@tempb \@tempa \fi \ifx
  \@tempb \@empty \def\@tempb {arXiv}\fi \@ifundefined
  {mn@eprint@\@tempb}{\@tempb:\@tempc}{\expandafter \expandafter \csname
  mn@eprint@\@tempb\endcsname \expandafter{\@tempc}}}

\bibitem[\protect\citeauthoryear{{Abbott} et~al.,}{{Abbott}
  et~al.}{2016a}]{2016PhRvX...6d1015A}
{Abbott} B.~P.,  et~al., 2016a, \mn@doi [Physical Review X]
  {10.1103/PhysRevX.6.041015}, \href
  {http://adsabs.harvard.edu/abs/2016PhRvX...6d1015A} {6, 041015}

\bibitem[\protect\citeauthoryear{{Abbott} et~al.,}{{Abbott}
  et~al.}{2016b}]{2016PhRvL.116f1102A}
{Abbott} B.~P.,  et~al., 2016b, \mn@doi [Physical Review Letters]
  {10.1103/PhysRevLett.116.061102}, \href
  {http://adsabs.harvard.edu/abs/2016PhRvL.116f1102A} {116, 061102}

\bibitem[\protect\citeauthoryear{{Abbott} et~al.,}{{Abbott}
  et~al.}{2016c}]{2016PhRvL.116x1103A}
{Abbott} B.~P.,  et~al., 2016c, \mn@doi [Physical Review Letters]
  {10.1103/PhysRevLett.116.241103}, \href
  {http://adsabs.harvard.edu/abs/2016PhRvL.116x1103A} {116, 241103}

\bibitem[\protect\citeauthoryear{{Abbott} et~al.,}{{Abbott}
  et~al.}{2017a}]{2017PhRvL.118v1101A}
{Abbott} B.~P.,  et~al., 2017a, \mn@doi [Physical Review Letters]
  {10.1103/PhysRevLett.118.221101}, \href
  {http://adsabs.harvard.edu/abs/2017PhRvL.118v1101A} {118, 221101}

\bibitem[\protect\citeauthoryear{{Abbott} et~al.,}{{Abbott}
  et~al.}{2017b}]{2017PhRvL.119n1101A}
{Abbott} B.~P.,  et~al., 2017b, \mn@doi [Physical Review Letters]
  {10.1103/PhysRevLett.119.141101}, \href
  {http://adsabs.harvard.edu/abs/2017PhRvL.119n1101A} {119, 141101}

\bibitem[\protect\citeauthoryear{{Abbott} et~al.,}{{Abbott}
  et~al.}{2017c}]{2017PhRvL.119p1101A}
{Abbott} B.~P.,  et~al., 2017c, \mn@doi [Physical Review Letters]
  {10.1103/PhysRevLett.119.161101}, \href
  {http://adsabs.harvard.edu/abs/2017PhRvL.119p1101A} {119, 161101}

\bibitem[\protect\citeauthoryear{{Baumgardt}, {Hopman}, {Portegies Zwart}  \&
  {Makino}}{{Baumgardt} et~al.}{2006}]{2006MNRAS.372..467B}
{Baumgardt} H.,  {Hopman} C.,  {Portegies Zwart} S.,   {Makino} J.,  2006,
  \mn@doi [\mnras] {10.1111/j.1365-2966.2006.10885.x}, \href
  {http://adsabs.harvard.edu/abs/2006MNRAS.372..467B} {372, 467}

\bibitem[\protect\citeauthoryear{{Blanchet}}{{Blanchet}}{2014}]{2014LRR....17....2B}
{Blanchet} L.,  2014, \mn@doi [Living Reviews in Relativity]
  {10.12942/lrr-2014-2}, \href
  {http://adsabs.harvard.edu/abs/2014LRR....17....2B} {17}

\bibitem[\protect\citeauthoryear{{Carter} \& {Luminet}}{{Carter} \&
  {Luminet}}{1985}]{1985MNRAS.212...23C}
{Carter} B.,  {Luminet} J.~P.,  1985, \mn@doi [\mnras]
  {10.1093/mnras/212.1.23}, \href
  {http://adsabs.harvard.edu/abs/1985MNRAS.212...23C} {212, 23}

\bibitem[\protect\citeauthoryear{{Clark}}{{Clark}}{1975}]{1975ApJ...199L.143C}
{Clark} G.~W.,  1975, \mn@doi [\apjl] {10.1086/181869}, \href
  {http://adsabs.harvard.edu/abs/1975ApJ...199L.143C} {199, L143}

\bibitem[\protect\citeauthoryear{{Diener}, {Kosovichev}, {Kotok}, {Novikov}  \&
  {Pethick}}{{Diener} et~al.}{1995}]{1995MNRAS.275..498D}
{Diener} P.,  {Kosovichev} A.~G.,  {Kotok} E.~V.,  {Novikov} I.~D.,   {Pethick}
  C.~J.,  1995, \mn@doi [\mnras] {10.1093/mnras/275.2.498}, \href
  {http://adsabs.harvard.edu/abs/1995MNRAS.275..498D} {275, 498}

\bibitem[\protect\citeauthoryear{{Fabian}, {Pringle}  \& {Rees}}{{Fabian}
  et~al.}{1975}]{1975MNRAS.172P..15F}
{Fabian} A.~C.,  {Pringle} J.~E.,   {Rees} M.~J.,  1975, \mn@doi [\mnras]
  {10.1093/mnras/172.1.15P}, \href
  {http://adsabs.harvard.edu/abs/1975MNRAS.172P..15F} {172, 15p}

\bibitem[\protect\citeauthoryear{{Fabrycky} \& {Tremaine}}{{Fabrycky} \&
  {Tremaine}}{2007}]{2007ApJ...669.1298F}
{Fabrycky} D.,  {Tremaine} S.,  2007, \mn@doi [\apj] {10.1086/521702}, \href
  {http://adsabs.harvard.edu/abs/2007ApJ...669.1298F} {669, 1298}

\bibitem[\protect\citeauthoryear{{Fregeau}, {Cheung}, {Portegies Zwart}  \&
  {Rasio}}{{Fregeau} et~al.}{2004}]{Fregeau2004}
{Fregeau} J.~M.,  {Cheung} P.,  {Portegies Zwart} S.~F.,   {Rasio} F.~A.,
  2004, \mn@doi [\mnras] {10.1111/j.1365-2966.2004.07914.x}, \href
  {http://adsabs.harvard.edu/abs/2004MNRAS.352....1F} {352, 1}

\bibitem[\protect\citeauthoryear{{Gaburov}, {Gualandris}  \& {Portegies
  Zwart}}{{Gaburov} et~al.}{2008}]{2008MNRAS.384..376G}
{Gaburov} E.,  {Gualandris} A.,   {Portegies Zwart} S.,  2008, \mn@doi [\mnras]
  {10.1111/j.1365-2966.2007.12731.x}, \href
  {http://adsabs.harvard.edu/abs/2008MNRAS.384..376G} {384, 376}

\bibitem[\protect\citeauthoryear{{Gaburov}, {Lombardi}  \& {Portegies
  Zwart}}{{Gaburov} et~al.}{2010}]{2010MNRAS.402..105G}
{Gaburov} E.,  {Lombardi} Jr. J.~C.,   {Portegies Zwart} S.,  2010, \mn@doi
  [\mnras] {10.1111/j.1365-2966.2009.15900.x}, \href
  {http://adsabs.harvard.edu/abs/2010MNRAS.402..105G} {402, 105}

\bibitem[\protect\citeauthoryear{{Gardner} et~al.,}{{Gardner}
  et~al.}{2006}]{2006SSRv..123..485G}
{Gardner} J.~P.,  et~al., 2006, \mn@doi [\ssr] {10.1007/s11214-006-8315-7},
  \href {http://adsabs.harvard.edu/abs/2006SSRv..123..485G} {123, 485}

\bibitem[\protect\citeauthoryear{{Giersz}}{{Giersz}}{1985a}]{1985AcA....35..119G}
{Giersz} M.,  1985a, \actaa, \href
  {http://adsabs.harvard.edu/abs/1985AcA....35..119G} {35, 119}

\bibitem[\protect\citeauthoryear{{Giersz}}{{Giersz}}{1985b}]{1985AcA....35..401G}
{Giersz} M.,  1985b, \actaa, \href
  {http://adsabs.harvard.edu/abs/1985AcA....35..401G} {35, 401}

\bibitem[\protect\citeauthoryear{{Giersz}}{{Giersz}}{1986}]{1986AcA....36..181G}
{Giersz} M.,  1986, \actaa, \href
  {http://esoads.eso.org/abs/1986AcA....36..181G} {36, 181}

\bibitem[\protect\citeauthoryear{{Giersz}, {Heggie}, {Hurley}  \&
  {Hypki}}{{Giersz} et~al.}{2013}]{Giersz2013}
{Giersz} M.,  {Heggie} D.~C.,  {Hurley} J.~R.,   {Hypki} A.,  2013, \mn@doi
  [\mnras] {10.1093/mnras/stt307}, \href
  {http://adsabs.harvard.edu/abs/2013MNRAS.431.2184G} {431, 2184}

\bibitem[\protect\citeauthoryear{G{\"u}ltekin, Miller  \&
  Hamilton}{G{\"u}ltekin et~al.}{2006}]{2006ApJ...640..156G}
G{\"u}ltekin K.,  Miller M.~C.,   Hamilton D.~P.,  2006, \apj, 640, 156

\bibitem[\protect\citeauthoryear{Hansen}{Hansen}{1972}]{Hansen:1972il}
Hansen R.,  1972, Phys. Rev. D, 5, 1021

\bibitem[\protect\citeauthoryear{{Hypki} \& {Giersz}}{{Hypki} \&
  {Giersz}}{2013a}]{Hypki2013}
{Hypki} A.,  {Giersz} M.,  2013a, \mn@doi [\mnras] {10.1093/mnras/sts415},
  \href {http://adsabs.harvard.edu/abs/2013MNRAS.429.1221H} {429, 1221}

\bibitem[\protect\citeauthoryear{{Hypki} \& {Giersz}}{{Hypki} \&
  {Giersz}}{2013b}]{2013MNRAS.429.1221H}
{Hypki} A.,  {Giersz} M.,  2013b, \mn@doi [\mnras] {10.1093/mnras/sts415},
  \href {http://adsabs.harvard.edu/abs/2013MNRAS.429.1221H} {429, 1221}

\bibitem[\protect\citeauthoryear{{Iben} \& {Livio}}{{Iben} \&
  {Livio}}{1993}]{1993PASP..105.1373I}
{Iben} Jr. I.,  {Livio} M.,  1993, \mn@doi [\pasp] {10.1086/133321}, \href
  {http://adsabs.harvard.edu/abs/1993PASP..105.1373I} {105, 1373}

\bibitem[\protect\citeauthoryear{{Ivanov} \& {Novikov}}{{Ivanov} \&
  {Novikov}}{2001}]{2001ApJ...549..467I}
{Ivanov} P.~B.,  {Novikov} I.~D.,  2001, \mn@doi [\apj] {10.1086/319050}, \href
  {http://adsabs.harvard.edu/abs/2001ApJ...549..467I} {549, 467}

\bibitem[\protect\citeauthoryear{{Ivanov}, {Chernyakova}  \&
  {Novikov}}{{Ivanov} et~al.}{2003}]{2003MNRAS.338..147I}
{Ivanov} P.~B.,  {Chernyakova} M.~A.,   {Novikov} I.~D.,  2003, \mn@doi
  [\mnras] {10.1046/j.1365-8711.2003.06028.x}, \href
  {http://adsabs.harvard.edu/abs/2003MNRAS.338..147I} {338, 147}

\bibitem[\protect\citeauthoryear{{Kochanek}}{{Kochanek}}{1992}]{1992ApJ...385..604K}
{Kochanek} C.~S.,  1992, \mn@doi [\apj] {10.1086/170966}, \href
  {http://adsabs.harvard.edu/abs/1992ApJ...385..604K} {385, 604}

\bibitem[\protect\citeauthoryear{{Kosovichev} \& {Novikov}}{{Kosovichev} \&
  {Novikov}}{1992}]{1992MNRAS.258..715K}
{Kosovichev} A.~G.,  {Novikov} I.~D.,  1992, \mn@doi [\mnras]
  {10.1093/mnras/258.4.715}, \href
  {http://adsabs.harvard.edu/abs/1992MNRAS.258..715K} {258, 715}

\bibitem[\protect\citeauthoryear{{Kozai}}{{Kozai}}{1962}]{1962AJ.....67..591K}
{Kozai} Y.,  1962, \mn@doi [\aj] {10.1086/108790}, \href
  {http://adsabs.harvard.edu/abs/1962AJ.....67..591K} {67, 591}

\bibitem[\protect\citeauthoryear{{Kupi}, {Amaro-Seoane}  \& {Spurzem}}{{Kupi}
  et~al.}{2006}]{2006MNRAS.371L..45K}
{Kupi} G.,  {Amaro-Seoane} P.,   {Spurzem} R.,  2006, \mn@doi [\mnras]
  {10.1111/j.1745-3933.2006.00205.x}, \href
  {http://adsabs.harvard.edu/abs/2006MNRAS.371L..45K} {371, L45}

\bibitem[\protect\citeauthoryear{{LSST Science Collaboration} et~al.,}{{LSST
  Science Collaboration} et~al.}{2009}]{2009arXiv0912.0201L}
{LSST Science Collaboration} et~al., 2009, preprint, \href
  {http://adsabs.harvard.edu/abs/2009arXiv0912.0201L} {} (\mn@eprint {arXiv}
  {0912.0201})

\bibitem[\protect\citeauthoryear{{Lai} \& {Shapiro}}{{Lai} \&
  {Shapiro}}{1995}]{1995ApJ...443..705L}
{Lai} D.,  {Shapiro} S.~L.,  1995, \mn@doi [\apj] {10.1086/175562}, \href
  {http://adsabs.harvard.edu/abs/1995ApJ...443..705L} {443, 705}

\bibitem[\protect\citeauthoryear{{Lai}, {Rasio}  \& {Shapiro}}{{Lai}
  et~al.}{1993a}]{1993ApJS...88..205L}
{Lai} D.,  {Rasio} F.~A.,   {Shapiro} S.~L.,  1993a, \mn@doi [\apjs]
  {10.1086/191822}, \href {http://adsabs.harvard.edu/abs/1993ApJS...88..205L}
  {88, 205}

\bibitem[\protect\citeauthoryear{{Lai}, {Rasio}  \& {Shapiro}}{{Lai}
  et~al.}{1993b}]{1993ApJ...406L..63L}
{Lai} D.,  {Rasio} F.~A.,   {Shapiro} S.~L.,  1993b, \mn@doi [\apjl]
  {10.1086/186787}, \href {http://adsabs.harvard.edu/abs/1993ApJ...406L..63L}
  {406, L63}

\bibitem[\protect\citeauthoryear{{Lai}, {Rasio}  \& {Shapiro}}{{Lai}
  et~al.}{1994a}]{1994ApJ...420..811L}
{Lai} D.,  {Rasio} F.~A.,   {Shapiro} S.~L.,  1994a, \mn@doi [\apj]
  {10.1086/173606}, \href {http://adsabs.harvard.edu/abs/1994ApJ...420..811L}
  {420, 811}

\bibitem[\protect\citeauthoryear{{Lai}, {Rasio}  \& {Shapiro}}{{Lai}
  et~al.}{1994b}]{1994ApJ...423..344L}
{Lai} D.,  {Rasio} F.~A.,   {Shapiro} S.~L.,  1994b, \mn@doi [\apj]
  {10.1086/173812}, \href {http://adsabs.harvard.edu/abs/1994ApJ...423..344L}
  {423, 344}

\bibitem[\protect\citeauthoryear{{Lee}}{{Lee}}{1993}]{1993ApJ...418..147L}
{Lee} M.~H.,  1993, \mn@doi [\apj] {10.1086/173378}, \href
  {http://adsabs.harvard.edu/abs/1993ApJ...418..147L} {418, 147}

\bibitem[\protect\citeauthoryear{{Lee} \& {Ostriker}}{{Lee} \&
  {Ostriker}}{1986}]{1986ApJ...310..176L}
{Lee} H.~M.,  {Ostriker} J.~P.,  1986, \mn@doi [\apj] {10.1086/164674}, \href
  {http://adsabs.harvard.edu/abs/1986ApJ...310..176L} {310, 176}

\bibitem[\protect\citeauthoryear{Luminet \& Carter}{Luminet \&
  Carter}{1986}]{Luminet:1986cha}
Luminet J.~P.,  Carter B.,  1986, \apjs, 61, 219

\bibitem[\protect\citeauthoryear{{MacLeod}, {Ostriker}  \& {Stone}}{{MacLeod}
  et~al.}{2018}]{2018arXiv180303261M}
{MacLeod} M.,  {Ostriker} E.~C.,   {Stone} J.~M.,  2018, preprint, \href
  {http://adsabs.harvard.edu/abs/2018arXiv180303261M} {} (\mn@eprint {arXiv}
  {1803.03261})

\bibitem[\protect\citeauthoryear{Mardling}{Mardling}{1995a}]{Mardling:1995hx}
Mardling R.~A.,  1995a, \apj, 450, 722

\bibitem[\protect\citeauthoryear{Mardling}{Mardling}{1995b}]{Mardling:1995it}
Mardling R.~A.,  1995b, \apj, 450, 732

\bibitem[\protect\citeauthoryear{Mardling \& Aarseth}{Mardling \&
  Aarseth}{2001}]{Mardling:2001dl}
Mardling R.~A.,  Aarseth S.~J.,  2001, \mnras, 321, 398

\bibitem[\protect\citeauthoryear{{Naoz}}{{Naoz}}{2016}]{2016ARA&A..54..441N}
{Naoz} S.,  2016, \mn@doi [\araa] {10.1146/annurev-astro-081915-023315}, \href
  {http://adsabs.harvard.edu/abs/2016ARA%26A..54..441N} {54, 441}

\bibitem[\protect\citeauthoryear{{Ogawaguchi} \& {Kojima}}{{Ogawaguchi} \&
  {Kojima}}{1996}]{1996PThPh..96..901O}
{Ogawaguchi} W.,  {Kojima} Y.,  1996, \mn@doi [Progress of Theoretical Physics]
  {10.1143/PTP.96.901}, \href
  {http://adsabs.harvard.edu/abs/1996PThPh..96..901O} {96, 901}

\bibitem[\protect\citeauthoryear{{Ogilvie}}{{Ogilvie}}{2014}]{2014ARA&A..52..171O}
{Ogilvie} G.~I.,  2014, \mn@doi [\araa] {10.1146/annurev-astro-081913-035941},
  \href {http://adsabs.harvard.edu/abs/2014ARA%26A..52..171O} {52, 171}

\bibitem[\protect\citeauthoryear{{Paczynski}}{{Paczynski}}{1976}]{1976IAUS...73...75P}
{Paczynski} B.,  1976, in {Eggleton} P.,  {Mitton} S.,   {Whelan} J.,  eds,
  IAU Symposium Vol. 73, Structure and Evolution of Close Binary Systems. p.~75

\bibitem[\protect\citeauthoryear{{Perets} \& {Naoz}}{{Perets} \&
  {Naoz}}{2009}]{2009ApJ...699L..17P}
{Perets} H.~B.,  {Naoz} S.,  2009, \mn@doi [\apjl]
  {10.1088/0004-637X/699/1/L17}, \href
  {http://adsabs.harvard.edu/abs/2009ApJ...699L..17P} {699, L17}

\bibitem[\protect\citeauthoryear{{Perets}, {Li}, {Lombardi}  \&
  {Milcarek}}{{Perets} et~al.}{2016}]{2016ApJ...823..113P}
{Perets} H.~B.,  {Li} Z.,  {Lombardi} Jr. J.~C.,   {Milcarek} Jr. S.~R.,  2016,
  \mn@doi [\apj] {10.3847/0004-637X/823/2/113}, \href
  {http://adsabs.harvard.edu/abs/2016ApJ...823..113P} {823, 113}

\bibitem[\protect\citeauthoryear{Peters}{Peters}{1964}]{Peters:1964bc}
Peters P.,  1964, Phys. Rev., 136, B1224

\bibitem[\protect\citeauthoryear{{Portegies Zwart} \& {Meinen}}{{Portegies
  Zwart} \& {Meinen}}{1993}]{1993A&A...280..174P}
{Portegies Zwart} S.~F.,  {Meinen} A.~T.,  1993, \aap, \href
  {http://adsabs.harvard.edu/abs/1993A%26A...280..174P} {280, 174}

\bibitem[\protect\citeauthoryear{Press \& Teukolsky}{Press \&
  Teukolsky}{1977}]{1977ApJ...213..183P}
Press W.~H.,  Teukolsky S.~A.,  1977, \apj, 213, 183

\bibitem[\protect\citeauthoryear{{Rodriguez}, {Amaro-Seoane}, {Chatterjee}  \&
  {Rasio}}{{Rodriguez} et~al.}{2017}]{2017arXiv171204937R}
{Rodriguez} C.~L.,  {Amaro-Seoane} P.,  {Chatterjee} S.,   {Rasio} F.~A.,
  2017, preprint, \href {http://adsabs.harvard.edu/abs/2017arXiv171204937R} {}
  (\mn@eprint {arXiv} {1712.04937})

\bibitem[\protect\citeauthoryear{{Samsing}}{{Samsing}}{2017}]{2017arXiv171107452S}
{Samsing} J.,  2017, preprint, \href
  {http://adsabs.harvard.edu/abs/2017arXiv171107452S} {} (\mn@eprint {arXiv}
  {1711.07452})

\bibitem[\protect\citeauthoryear{{Samsing} \& {Ramirez-Ruiz}}{{Samsing} \&
  {Ramirez-Ruiz}}{2017}]{2017ApJ...840L..14S}
{Samsing} J.,  {Ramirez-Ruiz} E.,  2017, \mn@doi [\apjl]
  {10.3847/2041-8213/aa6f0b}, \href
  {http://adsabs.harvard.edu/abs/2017ApJ...840L..14S} {840, L14}

\bibitem[\protect\citeauthoryear{{Samsing}, {MacLeod}  \&
  {Ramirez-Ruiz}}{{Samsing} et~al.}{2014}]{2014ApJ...784...71S}
{Samsing} J.,  {MacLeod} M.,   {Ramirez-Ruiz} E.,  2014, \mn@doi [\apj]
  {10.1088/0004-637X/784/1/71}, \href
  {http://adsabs.harvard.edu/abs/2014ApJ...784...71S} {784, 71}

\bibitem[\protect\citeauthoryear{{Samsing}, {Askar}  \& {Giersz}}{{Samsing}
  et~al.}{2017a}]{2017arXiv171206186S}
{Samsing} J.,  {Askar} A.,   {Giersz} M.,  2017a, preprint, \href
  {http://adsabs.harvard.edu/abs/2017arXiv171206186S} {} (\mn@eprint {arXiv}
  {1712.06186})

\bibitem[\protect\citeauthoryear{{Samsing}, {MacLeod}  \&
  {Ramirez-Ruiz}}{{Samsing} et~al.}{2017b}]{2017ApJ...846...36S}
{Samsing} J.,  {MacLeod} M.,   {Ramirez-Ruiz} E.,  2017b, \mn@doi [\apj]
  {10.3847/1538-4357/aa7e32}, \href
  {http://adsabs.harvard.edu/abs/2017ApJ...846...36S} {846, 36}

\bibitem[\protect\citeauthoryear{{Samsing}, {MacLeod}  \&
  {Ramirez-Ruiz}}{{Samsing} et~al.}{2018}]{2018ApJ...853..140S}
{Samsing} J.,  {MacLeod} M.,   {Ramirez-Ruiz} E.,  2018, \mn@doi [\apj]
  {10.3847/1538-4357/aaa715}, \href
  {http://adsabs.harvard.edu/abs/2018ApJ...853..140S} {853, 140}

\bibitem[\protect\citeauthoryear{{Spergel} et~al.,}{{Spergel}
  et~al.}{2013}]{2013arXiv1305.5422S}
{Spergel} D.,  et~al., 2013, preprint, \href
  {http://adsabs.harvard.edu/abs/2013arXiv1305.5422S} {} (\mn@eprint {arXiv}
  {1305.5422})

\bibitem[\protect\citeauthoryear{{Stone}, {K{\"u}pper}  \& {Ostriker}}{{Stone}
  et~al.}{2017}]{2017MNRAS.467.4180S}
{Stone} N.~C.,  {K{\"u}pper} A.~H.~W.,   {Ostriker} J.~P.,  2017, \mn@doi
  [\mnras] {10.1093/mnras/stx097}, \href
  {http://adsabs.harvard.edu/abs/2017MNRAS.467.4180S} {467, 4180}

\bibitem[\protect\citeauthoryear{{Taam} \& {Sandquist}}{{Taam} \&
  {Sandquist}}{2000}]{2000ARA&A..38..113T}
{Taam} R.~E.,  {Sandquist} E.~L.,  2000, \mn@doi [\araa]
  {10.1146/annurev.astro.38.1.113}, \href
  {http://adsabs.harvard.edu/abs/2000ARA%26A..38..113T} {38, 113}

\bibitem[\protect\citeauthoryear{{Trani}, {Mapelli}, {Spera}  \&
  {Bressan}}{{Trani} et~al.}{2016}]{2016ApJ...831...61T}
{Trani} A.~A.,  {Mapelli} M.,  {Spera} M.,   {Bressan} A.,  2016, \mn@doi
  [\apj] {10.3847/0004-637X/831/1/61}, \href
  {http://adsabs.harvard.edu/abs/2016ApJ...831...61T} {831, 61}

\bibitem[\protect\citeauthoryear{{Turner}}{{Turner}}{1977}]{1977ApJ...216..610T}
{Turner} M.,  1977, \mn@doi [\apj] {10.1086/155501}, \href
  {http://adsabs.harvard.edu/abs/1977ApJ...216..610T} {216, 610}

\bibitem[\protect\citeauthoryear{{Tylenda} et~al.,}{{Tylenda}
  et~al.}{2011}]{2011A&A...528A.114T}
{Tylenda} R.,  et~al., 2011, \mn@doi [\aap] {10.1051/0004-6361/201016221},
  \href {http://adsabs.harvard.edu/abs/2011A%26A...528A.114T} {528, A114}

\bibitem[\protect\citeauthoryear{{Vick} \& {Lai}}{{Vick} \&
  {Lai}}{2018}]{2018MNRAS.476..482V}
{Vick} M.,  {Lai} D.,  2018, \mn@doi [\mnras] {10.1093/mnras/sty225}, \href
  {http://adsabs.harvard.edu/abs/2018MNRAS.476..482V} {476, 482}

\makeatother
\end{thebibliography}


\bsp	
\label{lastpage}
\end{document}